\documentclass[11pt]{article}

\usepackage{times}
\usepackage{soul}
\usepackage{url}
\usepackage[hidelinks]{hyperref}
\usepackage[utf8]{inputenc}
\usepackage[small]{caption}
\usepackage{graphicx}
\usepackage{amsmath}
\usepackage{amsthm}
\usepackage{booktabs}
\usepackage{algorithm}
\usepackage{algorithmic}
\usepackage[switch]{lineno}
\usepackage{xcolor}
\usepackage{mathtools}
\usepackage{amssymb}
\usepackage{multirow}
\usepackage{multicol}
\usepackage{comment}  

\newtheorem{definition}{Definition}

\title{Deliberation and Voting in Approval-Based Multi-Winner Elections}

\author{
Kanav Mehra$^1$ \and
Nanda Kishore Sreenivas$^1$\and
Kate Larson$^1$
}

\date{%
    $^1$David R. Cheriton School of Computer Science, University of Waterloo \\
    \{kanav.mehra, nksreenivas, kate.larson\}@uwaterloo.ca
}
\begin{document}

\maketitle

\begin{abstract}
Citizen-focused democratic processes where participants deliberate on alternatives and then vote to make the final decision are increasingly popular today. While the computational social choice literature has extensively investigated voting rules, there is limited work that explicitly looks at the interplay of the deliberative process and voting. In this paper, we build a deliberation model using established models from the opinion-dynamics literature and study the effect of different deliberation mechanisms on voting outcomes achieved when using well-studied voting rules. Our results show that deliberation generally improves welfare and representation guarantees, but the results are sensitive to how the deliberation process is organized. We also show, experimentally, that simple voting rules, such as approval voting, perform as well as more sophisticated rules such as proportional approval voting \cite{thiele1895om} or method of equal shares \cite{rule-x,peters2021proportional} if deliberation is properly supported. This has ramifications on the practical use of such voting rules in citizen-focused democratic processes.
\end{abstract}

\section{Introduction}
Scenarios, where a committee must be selected to represent the interests of some larger group, are ubiquitous, ranging from political domains such as parliamentary elections and  participatory budgeting (PB) \cite{cabannes2004participatory} to technical applications such as designing recommender systems \cite{SKOWRON2016191} and diversifying search results \cite{proportionalijcai2017p58}.
\emph{Multi-winner} voting has been well studied within the social choice literature, with a focus on understanding how the ``best" committee can be selected. However, even defining what is meant by ``best" is no trivial undertaking.  In some contexts, such as aggregation of expert judgements, one might want a committee that consists of the highest-rated $k$ alternatives. However, in other tasks such as choosing $k$ locations for constructing a public facility (\textit{e.g.} hospitals, fire stations), it is preferable to ensure as many  voters as possible have access to the facility.

In citizen-focused democratic processes such as citizens' assemblies \cite{scotland_2022} and participatory budgeting \cite{cabannes2004participatory}, there exists extensive scope for discussion over the multitude of possible alternatives. For example, deliberation is an important phase in most implementations of participatory budgeting as it allows voters to refine their preferences and facilitates the exchange of information, with the objective of reaching consensus \cite{nisarg2021participatory}. While deliberation is a vital component of democratic processes \cite{fishkin2009people,Habermas1996-HABBFA-2}, it cannot completely replace voting because, in reality, deliberation does not guarantee unanimity. A decision must still be made. 
Accordingly, we argue that it is essential to understand the relationship between voting and deliberation.  To this end, we 
  (1) introduce an agent-based model of deliberation, and 
  (2)  study the effect of different deliberation mechanisms on the outcomes obtained when using well-studied voting rules.
  

We focus on approval-based elections, where 
 voters  express  preferences by sharing a subset of approved candidates. Approval ballots are used in practice due to their simplicity and flexibility \cite{brams2007approval,benade2021preference,nisarg2021participatory}. They also offer scope for deliberation as often voters are left to decide between many different alternatives. 
 We present an agent-based model of deliberation and explore various alternatives for structuring deliberation groups. We evaluate standard multi-winner voting rules, both before and after voters have the opportunity to deliberate, with respect to standard objectives from the literature, including social welfare, representation, and proportionality.
We show that deliberation, in almost all scenarios, significantly improves welfare, representation, and proportionality. However, the results are sensitive to the deliberation mechanism; increased exposure to diverse opinions (or agents from different backgrounds) leads to better outcomes. More importantly, our results indicate that in the presence of effective deliberation, \emph{simple}, explainable voting rules such as approval voting  perform as well as more sophisticated, \emph{complex} rules. This can serve to guide the design and deployment of voting rules in citizen-focused democratic processes.

\noindent{\bf Related Work: }
The social choice literature has extensively studied the quality of approval-based multi-winner voting rules. From the quantitative perspective, a recent paper \cite{lackner_2020} provides an in-depth theoretical and empirical analysis of different approval-based multi-winner voting rules with respect to (utilitarian) social welfare and representation guarantees. \cite{welfarevsrep} extended this work to study the welfare-representation trade-off in the more general PB setting. The traditional axiomatic approach, on the other hand, provides a qualitative evaluation, \emph{i.e.} whether a voting rule satisfies a property or not. For approval-based rules, recent work has focused heavily on proportionality axioms \cite{aziz2017justified,sanchez2017proportional,aziz2018complexity,brill2017phragmen,lackner2018consistent,skowron2021proportionality}. We refer the reader to \cite{faliszewski2017multiwinner} for an extensive survey on the properties of multi-winner rules.

Deliberation, specifically within social choice, has been studied through multiple approaches. From the theoretical perspective, a wide variety of mathematical deliberation models have been proposed \cite{chung2020formal,zvi2021iterative}. For example, \cite{goel2016towards,fain2017sequential} introduce iterative small-group deliberation mechanisms for reaching consensus in collective decision-making problems. \cite{elkind2021united} propose a consensus-reaching deliberation protocol based on coalition formation. A recent experimental study \cite{rad2021deliberation} shows that deliberation leads to meta-agreements and single-peaked preferences under specific conditions. \cite{perote2015model} look at deliberation and voting simultaneously, but their work is limited to the \emph{ground-truth} setting with ordinal preferences over three alternatives. They do not study the impact of deliberation on the quantitative and qualitative properties of voting rules.

In this paper, we bridge the gap between deliberation and voting literature. To our knowledge, we are the first to experimentally study the effect of deliberation on voting outcomes across different deliberation strategies.

\section{Preliminaries}
Let $E = (C,N)$ be an election, where $C = \{c_1, c_2,..., c_m\}$ and $N = \{1,...,n\}$ are sets of $m$ candidates and $n$ voters, respectively. Each voter $i \in N$, has an \textit{approval ballot} $A_i \subseteq C$, containing the set of its approved candidates. The \textit{approval profile} $A = \{A_1, A_2,..., A_n\}$ represents the approval ballots for all voters. For a candidate $c_j \in C$, $N(c_j)$ is the set of voters that approve $c_j$ and its \textit{approval score}, $V(c_j) = |N(c_j)|$. Let $S_k(C)$ denote all $k$-sized subsets of the candidate set $C$, representing the set of all possible committees of size $k$. Given approval profile $A$ and desired committee size $k \in \mathbb{N}$, the objective of a multi-winner election is to select a subset of candidates that form the winning committee $W\in S_k(C)$. An \textit{approval-based committee rule}, $R (A,k)$, is a social choice function that takes as input an approval profile $A$ and committee size $k$ and returns a set of \textit{winning committees}.\footnote{A tie-breaking method is used to pick one winning committee in cases where multiple winning committees are returned.} For any voting rule $R(A,k)$, we will use $W_R$ to denote its selected committee (after tie-breaking). 

\subsection{Properties}
\label{properties}

We ideally want our voting rules to exhibit certain desired properties, representing the principles that should govern the selection of winners given individual ballots. In this paper we compare voting rules across three dimensions: \textit{social welfare}, \textit{representation}, and \textit{proportionality}. Intuitively, the \textit{welfare} objective focuses on selecting candidates that garner maximum support from the voters.
\textit{Representation} cares about \textit{diversity}; carefully selecting a committee that maximizes the number of voters represented in the winning committee. A voter is represented if the final committee contains at least one of its approved candidates.

\begin{definition}
    [Utilitarian Social Welfare]
    For a given approval profile $A$ and committee size $k$, the utilitarian social welfare of a committee $W$ 
    is:
    \begin{align}
    \label{eq:1}
    SW(A, W) = \sum_{i \in N} \sum_{c \in W} u_i(c),    
    \end{align}
     $u_i(c) \in \mathbb{R}$ is the utility voter $i$ derives from candidate $c$.
\end{definition}

\begin{definition}
    [Representation Score]
    For a given approval profile $A$ and committee size $k$, the representation score of a committee $W$ is defined as:
    \begin{align}
    \label{eq:2}
    RP(A, W) = \sum_{i \in N} \min(1, |A_i \cap W|)    
    \end{align}
\end{definition}




It may not be possible to maximize both social welfare and representation, so 
\textit{proportionality} serves as an important third objective to  capture a compromise between welfare and representation.  It requires that if a large enough voter group collectively approves a shared candidate set, then the group must be ``fairly represented''.  Definitions of proportionality differ based on how they interpret ``fairly represented''.

\begin{definition}
[T-Cohesive Groups]
Consider an election $E = (C,N)$ with $n$ voters and committee size $k$. For any integer $T \ge 1$, a group of voters $N'$ is \textit{T-cohesive} if it contains at least $Tn/k$ voters and collectively approves at least $T$ common candidates, i.e. if $|\cap_{i \in N'} A_i| \ge T$ and $|N'| \ge Tn/k$.
\end{definition}
\begin{definition}
[Proportional Justified Representation (PJR)]
A committee $W$ of size $k$ satisfies PJR if for each integer $T \in \{1,...,k\}$ and every \textit{T-cohesive} group $N' \subseteq N$, it holds that $|(\cup_{i \in N'} A_i) \cap W| \ge T$.
\end{definition}
\begin{definition}
[Extended Justified Representation (EJR)]
A committee $W$ of size $k$ satisfies EJR if for each integer $T \in \{1,...,k\}$, every \textit{T-cohesive} group $N' \subseteq N$ contains at least one voter that approves at least \textit{T} candidates in W, i.e. for some $i \in N'$, $|A_i \cap W| \ge T$.
\end{definition}
Unlike EJR \cite{aziz2017justified}, where the focus is on a single group member, PJR provides a more natural requirement for group representation. However, EJR provides stronger guarantees for average voter satisfaction and implies PJR \cite{sanchez2017proportional}.


\subsection{Multi-winner Voting Rules}
\label{rules}
In this section, we introduce the set of approval-based multi-winner voting rules that form the basis of our analysis.

\noindent{\textbf{Approval Voting (AV):}} Given approval profile $A$ and a committee $W$, the AV-score is $sc_{av}(A,W) = \sum_{c \in W} V(c)$.
The AV rule is defined as  $R_{AV}(A,k) = \\ \arg\max_{W \in S_k(C)} sc_{av}(A, W).$ 
This rule selects $k$ candidates with the highest individual approval scores.

\noindent \textbf{Approval Chamberlin-Courant (CC):} The CC rule \cite{chamberlin1983representative}, $R_{CC}(A,k)$, picks committees that maximize representation score $RP(A,W)$. Given profile $A$, $R_{CC}(A,k) = \arg\max_{W \in S_k(C)} RP(A, W).$ 
It maximizes voter coverage by maximizing the number of voters with at least one approved candidate in the winning committee.

\noindent\textbf{Proportional Approval Voting (PAV):} \cite{thiele1895om} For profile $A$ and committee $W$, the PAV-score is defined as $sc_{pav}(A,W) = \sum_{i \in N} h(|W \cap A_i|)$, where $h(t) = \sum_{i=1}^{t} 1/i.$ The PAV rule is defined as $R_{PAV}(A,k)=\arg\max_{W \in S_k(C)} sc_{pav}(A, W)$. Based on the idea of diminishing returns, a voter's utility from having an approved candidate in the elected committee $W$ decreases according to the harmonic function $h(t)$. It is a variation of the AV rule that ensures proportional representation, as it guarantees EJR \cite{aziz2017justified}. PAV is the same as AV when committee size $k=1$, but computing PAV is NP-hard \cite{aziz2014computational}.

\noindent\textbf{Method-of-Equal-Shares (MES):} 
$R_{MES}(A,k)$, also known in the literature as Rule-X \cite{peters2021proportional,rule-x}, is an iterative process that uses the idea of budgets to guarantee proportionality. Each voter starts with a budget of $k/n$ and each candidate is of unit cost. In round $t$, a candidate $c$ is added to $W$ if it is $q$-affordable, \textit{i.e.} for some $q \geq 0, \sum_{i \in N(c)} \min(q, b_i(t)) \geq 1$, where $b_i(t)$ is the budget of voter $i$ in round $t$. If a candidate is successfully added then the budget of each supporting voter is reduced accordingly. This process continues until either $k$ candidates are added to the committee or it fails (then another voting rule is used to select the remaining candidates).

We elect to study these rules since they exhibit a wide range of properties, allowing for comparisons to be drawn across several axes. 
First, AV is known to maximize social welfare under certain conditions on voters' utility functions~\cite{lackner_2020,lackner2018consistent}, however, there are no guarantees that AV satisfies proportionality (EJR criterion) ~\cite{aziz2017justified}. On the other hand, CC maximizes representation, but its welfare properties are less well understood. 
Both PAV and MES guarantee EJR and maintain a balance between representation and social welfare. Thus, this collection of multi-winner voting rules covers the set of properties we are interested in better understanding.

\subsection{Objectives}
\label{sec:objectives}

We compare the above  voting rules according to different standard objectives \cite{lackner_2020}. In particular, we consider objectives across three dimensions: welfare, representation, and proportionality.

\noindent\textbf{Utilitarian Ratio:} This ratio compares the (utilitarian) social welfare achieved by $W_R=R(A,k)$ to the maximum social welfare achievable:
\begin{align}
        UR(R) = \dfrac{SW(A, W_R)}{max_{W \in S_k(C)} SW(A,W)}
    \end{align}

\noindent\textbf{Representation Ratio:}   This ratio measures the diversity of the committee $W_R=R(A,k)$, by comparing the representation score achieved by $W_R$ to the optimal representation score amongst all $k$-sized committees:
\begin{align}
        RR(R) = \dfrac{RP(A, W_R)}{max_{W \in S_k(C)} RP(A, W)}.
    \end{align}
Note that $RR(R_{CC})=1$.

\noindent\textbf{Utility-Representation Aggregate Score:} 
This score captures how well a voting rule, $R(A,k)$ balances both social welfare and representation:
\begin{align}
        URagg(R) = UR(R)*RR(R)
    \end{align}

Finally, we are interested in experimentally verifying whether or not the generated profile instances {\bf satisfy EJR or PJR}. To this end, we count the number of profile instances that satisfy these two properties.
\section{Deliberation Models}
We describe how we model deliberation processes. We first define our underlying agent population and how we model their initial preferences. We then discuss the deliberation process, through which agents exchange information and update their preferences. Finally, we observe that deliberation is often done, not at the full population level, but instead in smaller subgroups. We discuss different ways these deliberation subgroups can be created.

\subsection{Voting Population: Preferences and Utilities}
\label{sec:prefs}

Our agent population $N$ is divided into two sets --- a \emph{majority} and \emph{minority}, where the number of agents in the majority is greater than that in the minority. Agents' initial preferences depend on their population group. Consistent with previous work \cite{lackner_2020}, our preference model is based on the ordinal Mallows model. The rankings are then converted to an approval ballot using the top-ranked candidates. In particular, we assume an agent $i$'s initial preference ranking, $P_i^0$, is sampled from a Mallows model~\cite{mallows_1957}, with reference rankings, $\Pi_\mathrm{maj}$ and $\Pi_\mathrm{min}$, for the majority and minority populations respectively.\footnote{The Mallows model is a standard noise model for preferences. It  defines a probability distribution over rankings over alternatives (i.e. preferences), defined as  
$ \mathbb{P}(r) = \frac{1}{Z} \phi^{d(r,\Pi)}$ where $\Pi$ is a reference ranking, $d(r,\Pi)$ is the Kendall-tau distance between $r$ and $\Pi$, and $Z$ is a normalizing factor.}

We further assume that agents have underlying cardinal utilities for candidates, consistent with their ordinal preferences. 
These are represented by a vector $U_i = \langle u_i(c_1), u_i(c_2), \hdots, u_i(c_m) \rangle$, where $u_i(c_x)\geq u_i(c_y)$ if and only if $c_x\succeq_i c_y$ in $P_i^0$, and $u_i(c_x)\in[0,1]$. We work in this cardinal space as it allows us to leverage standard deliberation models and measure welfare across voting rules in settings where voters derive some utility from elected candidates who were not on their ballot. Our particular instantiation of utility functions subsumes earlier work (e.g.~\cite{lackner_2020}) and is consistent with utility models used in the social choice literature (e.g.~\cite{procaccia2006distortion,fain_2018}).

\subsection{The Deliberation Process}\label{sec:del_proc}

Deliberation is defined as a ``discussion in which individuals are amenable to scrutinizing and changing their preferences in the light of persuasion (but not manipulation, deception or coercion) from other participants"~\cite{dryzek_2003}. Deliberation thus requires a group of peers with whom to deliberate and a methodology for changing preferences. In this section we describe the process in which agents update their preferences, deferring  details about peer groups until later.\footnote{As is common in much of the deliberation literature (e.g \cite{dryzek_2003,perote2015model}), we assume agents are non-strategic and truthfully reveal their utilities.}

Deliberation is an iterative process, involving, at each step, a speaker and listeners. The speaker makes a report, based on their preferences, and the listeners update their own preferences based on this information. In this work, we use a variation of the Bounded Confidence (BC) model to capture the (abstract) deliberation process~\cite{hegselmann_2002}. The BC model is a particularly good match for modelling deliberation in groups because it was intended to ``describe formal meetings, where there is an effective interaction involving many people at the same time"~\cite{castellano_2009}.  In the BC model, listeners consider the speaker's report (e.g. utilities for different candidates), and update their opinions/preferences of the candidates independently,   only if the speaker's report is not ``too far'' from their own. The notion of distance is captured by a confidence parameter for each listener,  $\Delta_i$, where agents may have different confidence levels~\cite{lorenz_2007,weisbuch2002meet}. We make a simplifying assumption that agents' utilities for all $m$ candidates are independent of each other, and apply the BC model to each dimension (candidate) independently. 
The interested reader is referred to Appendix~\ref{appendix:op_dyn} for more details on the original BC model.

Given time step $t$, some agent, $x$, selected as the speaker, makes its report (which reveals $x$'s thoughts and utilities for the candidates). Each listener updates its own preferences across candidates $c_j\in C$ according to the following rule
\begin{equation}
    \resizebox{\linewidth}{!}{$
    u_i^{t+1}(c_j) = 
    \begin{cases}
        (1 - w_{ix}) u_i^t(c_j) + w_{ix} u_x^t(c_j), & \text{if } |u_i^t(c_j) - u_x^t(c_j)| \leq \Delta_i \\
        u_i^t(c_j), & \text{otherwise} 
    \end{cases}
    \label{eq:util_update}$}
\end{equation}
where $w_{ix}\in[0,1]$ is the \emph{influence weight} that $i$ places on $x$'s perspective. 
It is known that opinions from sources similar to oneself have a higher influence than opinions from dissimilar sources~\cite{wilder_1990,Mackie_1990}. To capture this phenomenon, we let $w_{ix}$ take on one of two values, contingent on the relationship between $i$ and $x$. If $i$ and $x$ are both members of the majority group ($N_{maj}$) or the minority group ($N_{min}$) then $w_{ix}=\alpha_i$, otherwise $w_{ix}=\beta_i$ where $\alpha_i\geq\beta_i$.

\subsection{Deliberation Groups}
\label{sec:groups}
In the real world, deliberation typically happens in small discussion or peer groups~\cite {scotland_2022,golz_2022}.  To this end, we divide the agent population into $g$ sub-groups of approximately equal size. The deliberation process is conducted within these sub-groups where one \emph{round} of deliberation is complete when all agents in each  group have had the opportunity to speak.

We are interested in understanding how group-formation strategies influence the deliberation process and the final decision made through voting.   Our strategies are informed by common heuristics or rationale used in practice and none rely on private/unknown information such as the agents' underlying utilities or preferences. 
We do, however, assume that whether an agent is a member of the majority or minority group is public information and allow  group-formation strategies to use such information. Finally, we consider both single-round and iterative group-formation strategies where  agents are divided into different groups in each round~\cite{scotland_2022}.

\subsubsection{Single-Round  Group-Formation Strategies}

\noindent\textbf{Homogeneous group:} Each group contains only agents who are members of $N_{maj}$ or $N_{min}$. That is, there is no mixing of minority and majority agents.

\noindent\textbf{Heterogeneous group:} Each group is selected such that the ratio of the number of majority agents to the number of minority agents within the group is  approximately equal to the majority:minority ratio in the overall population. Each group created through this strategy is   \emph{diverse} and representative of the overall agent population. This strategy is already popular among practitioners in the real world. Citizens' Assembly of Scotland diversifies deliberation groups based on age, gender, and political affiliation~\cite{golz_2022}. 

\noindent\textbf{Random group:}  Each group is created by  randomly sampling agents  from the population (without replacement) with equal probability. 

\noindent\textbf{Large group:}  This is a special case where the deliberation process runs over the entire population of agents.
 Considering  time constraints, limited attention spans, and other physical limitations, such a strategy is not typically used in practice. However, we include this strategy as a Utopian baseline because it ensures maximum exposure to the preferences of every other agent in the system.

\subsubsection{Iterative Group-Formation Strategies}

\noindent\textbf{Iterative random:} In each round, agents are randomly assigned to groups. 

\noindent\textbf{Iterative golfer: } This strategy is a variant of the social golfer problem~\cite{harvey_2002,liu_2019} from combinatorial optimization. The number of rounds, $R$, is fixed \emph{a priori}, and the number of times any pair of agents meet more than once is minimized. We refer the reader to Appendix~\ref{appendix:iter_golfer} for details.  A similar approach is used in Sortition Foundation's GroupSelect algorithm~\cite{verpoort_2020}, which is used by several nonprofits for group-formation in PB sessions~\cite{golz_2022}. 

\section{Experimental Setup}
\label{sec:exp_setup}
Our election setup consists of 50 candidates ($|C|=50$)\footnote{Typically, project proposals are invited from the participants in PB~\cite{cabannes2004participatory,nisarg2021participatory}. So, there are a large number of candidate projects to choose from (e.g., PB instances in Warsaw, Poland had between 20-100 projects (36 on average).\cite{stolicki2020pabulib,welfarevsrep}).} and 100 voters, with 80 agents in the majority group ($N_{maj}$) and 20 in the minority group ($N_{min}$).
Agents' initial preferences are sampled using a Mallows model, with $\phi=0.2$.  
Reference rankings, $\Pi_{maj}$ and $\Pi_{min}$, are sampled uniformly from all linear orders over $C$.
Due to this sampling process, agents in either the majority or minority group have fairly similar preferences (as $\phi$ is relatively small) but the two groups themselves are distinct. To instantiate agents' utility functions, we  generate $m$ samples independently from the uniform distribution $\mathbf{U}(0,1)$, sort it, and then map the utilities to the candidates according to the agent's preference ranking. For the BC model, all three parameters ($\Delta_i, \alpha_i, \beta_i$) are sampled from uniform distributions over the full range for each parameter.\footnote{We  ran experiments where all parameters were drawn from a normal distribution. There were no significant differences from the results reported here. }

When deliberating, agents are divided into 10 groups (except for the \emph{large group} strategy). This is similar to the Citizens’ Assembly of Scotland, which ran over 16 sessions; in each session, the 104 participants were divided across 12 tables~\cite{golz_2022}. For iterative deliberation, the deliberation continues for $R = 5$ rounds.


We consider different approval-based multi-winner voting rules to elect $k=5$ winners. We use a flexible ballot size, such that each agent's ballot is of size $b_i$, where $b_i$ is sampled from $\mathcal{N}(2k, 1.0)$.
 Agent $i$'s approval vote is then the set consisting of its $b_i$ top-ranked candidates from its preference ranking $P_i$.

As a baseline, we apply every voting rule to the agent preferences \emph{before} deliberation. We then run the different deliberation strategies, freezing agents' utilities once deliberation has concluded. We then apply every voting rule to the updated preferences. We use the Python library \textit{abcvoting}~\cite{abcvoting_2021}, and use random tie-breaking when a voting rule returns multiple winning committees.\footnote{The code for the experiments is available at: \url{https://github.com/kanav-mehra/deliberation-voting}}

To avoid trivial profiles, \emph{i.e.}, profiles where an almost perfect compromise between welfare and representation is easily achievable, we impose some eligibility conditions. An initial approval profile $A^0$ is eligible only if $ \text{RR}(AV, A^0) < 0.9 \land \text{UR}(CC, A^0) < 0.9$. This is a common technique used in simulations comparing voting rules based on synthetic datasets~\cite{lackner_2020}. 

This entire simulation is repeated $10,000$ times and the average values are reported. To determine statistical significance while comparing any two sets of results, we used both the $t$-test and Wilcoxon signed-rank test, and we found the $p$-values to be roughly similar. All pairs of comparisons between deliberation group strategies for a given voting rule are statistically significant ($p < 0.05$) unless otherwise noted.

\section{Results}

\begin{figure}[t]
    \centering
    \includegraphics[width=\linewidth]{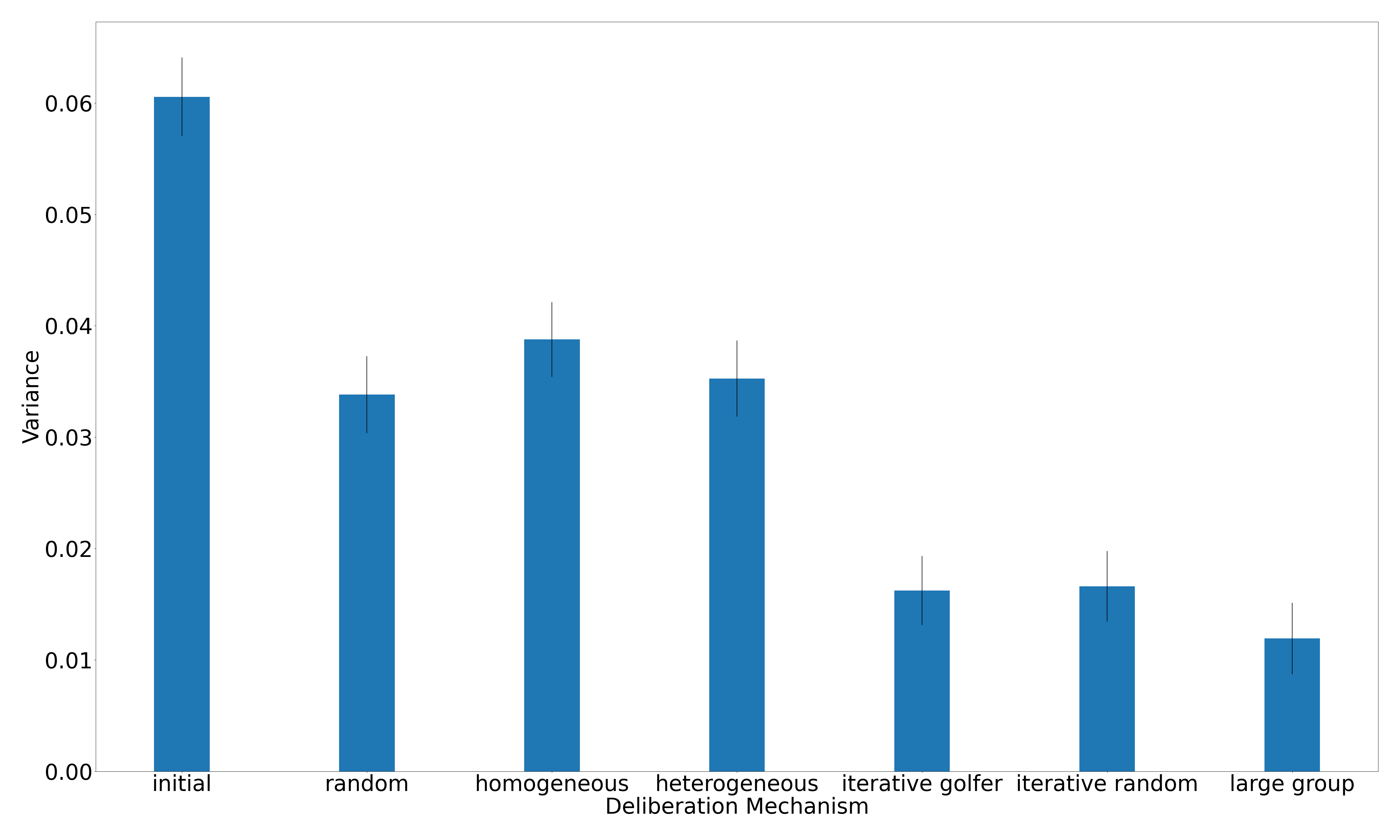}
    \caption{Average variance of agents' utilities for candidates. Lower variance implies a higher degree of consensus in the population.}
    \label{fig:variance}
\end{figure}
\noindent{\bf Impact of Deliberation on Preferences:}
To understand how deliberation processes shape and change agents' preferences, we compare the average variance in the agents' utilities before (\emph{initial}) and after deliberation (Figure~\ref{fig:variance}). As expected, deliberation reduces disagreement amongst agents, moving all towards a consensus. Processes where agents are exposed to more, diverse, agents (e.g. the iterative variants and the \emph{large group}) see the largest reduction in  variance across the population. We also measure consensus as the average number of common approvals between majority and minority voters and observe the same trend (see Appendix \ref{appendix:ballot_disagreement}). While achieving greater agreement is desirable, it should not be achieved by disregarding initial minority opinions. We delve into this topic in Section \ref{sec:discussion}.

\noindent\textbf{Utilitarian Ratio: }
\begin{figure}[t]
    \centering
    \includegraphics[width=\linewidth]{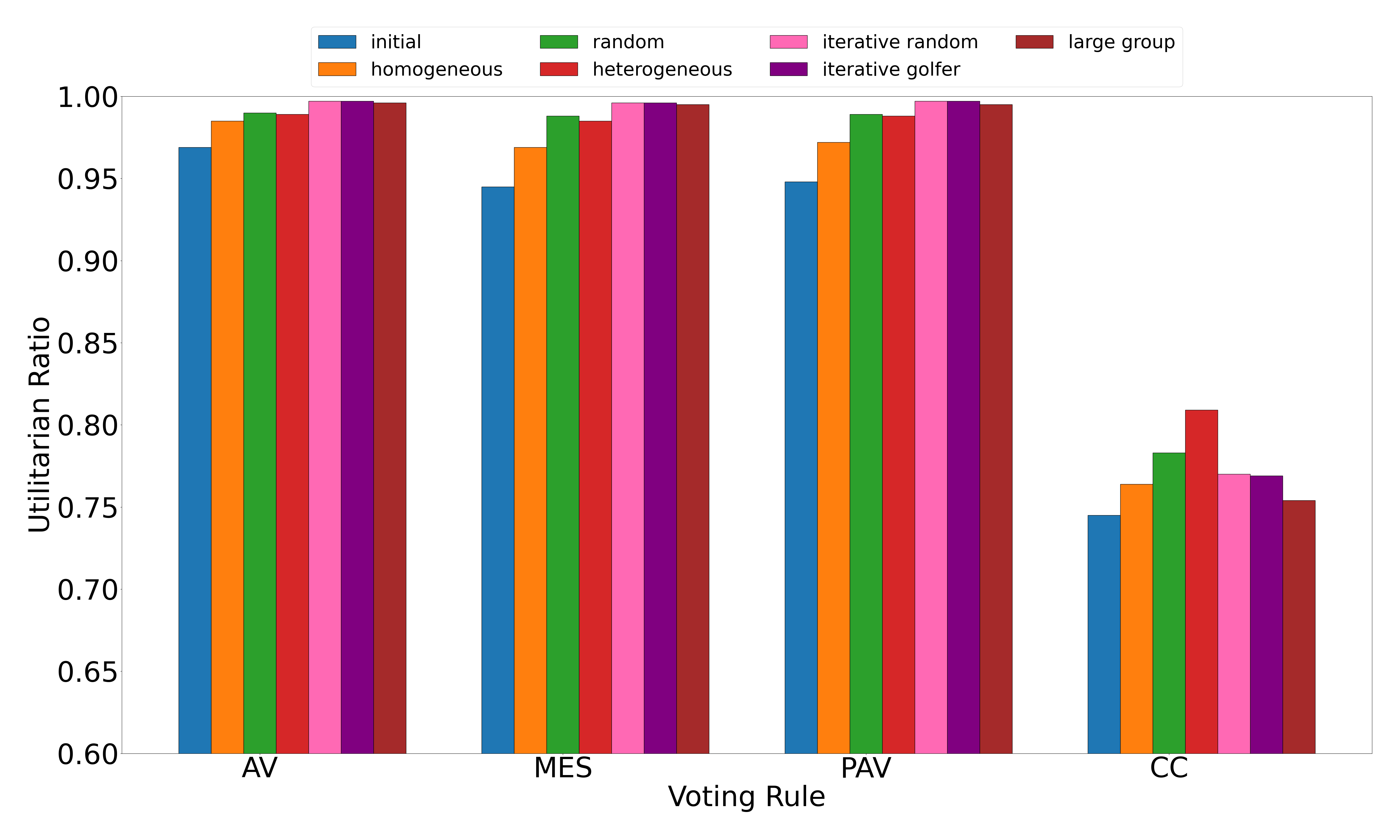}
    \caption{Utilitarian ratio across deliberation mechanisms. }
    \label{fig:utilitarian_ratio}
\end{figure}
Figure~\ref{fig:utilitarian_ratio} reports the impact of deliberation on utilitarian social welfare. 
First, we compare the voting rules where there is no deliberation (see \textit{initial} case denoted by the blue bars).
AV achieves the highest utilitarian ratio, \emph{i.e.} the utilitarian social welfare provided by AV is closest to the optimal social welfare. Both proportional rules (MES and PAV) are similar and obtain utilitarian ratios that are only slightly lower than AV. Finally, CC performs the worst in terms of welfare. In general, our results match the trends reported in previous work \cite{lackner_2020}.

We now  address our  main point of interest -- the effect of deliberation. As seen in Figure \ref{fig:utilitarian_ratio}, deliberation improves social welfare over the \textit{initial} baseline (blue). In single-round deliberation, both \textit{random} (green) and \textit{heterogeneous} (red) methods show similar results and outperform \textit{homogeneous} (orange) for AV, MES, and PAV. Iterative deliberation exhibits further improvement for these rules.  It is worth noting that \textit{iterative golfer} (purple) and \textit{iterative random} (pink) perform similarly and match the \textit{large group} benchmark (brown). For CC, social welfare is always improved with deliberation, however, iterative deliberation is not as powerful. This is due to the nature of the rule (explained in detail in Appendix \ref{appendix:iter_cc}).

\begin{figure}[t]
    \centering
    \includegraphics[width=\linewidth]{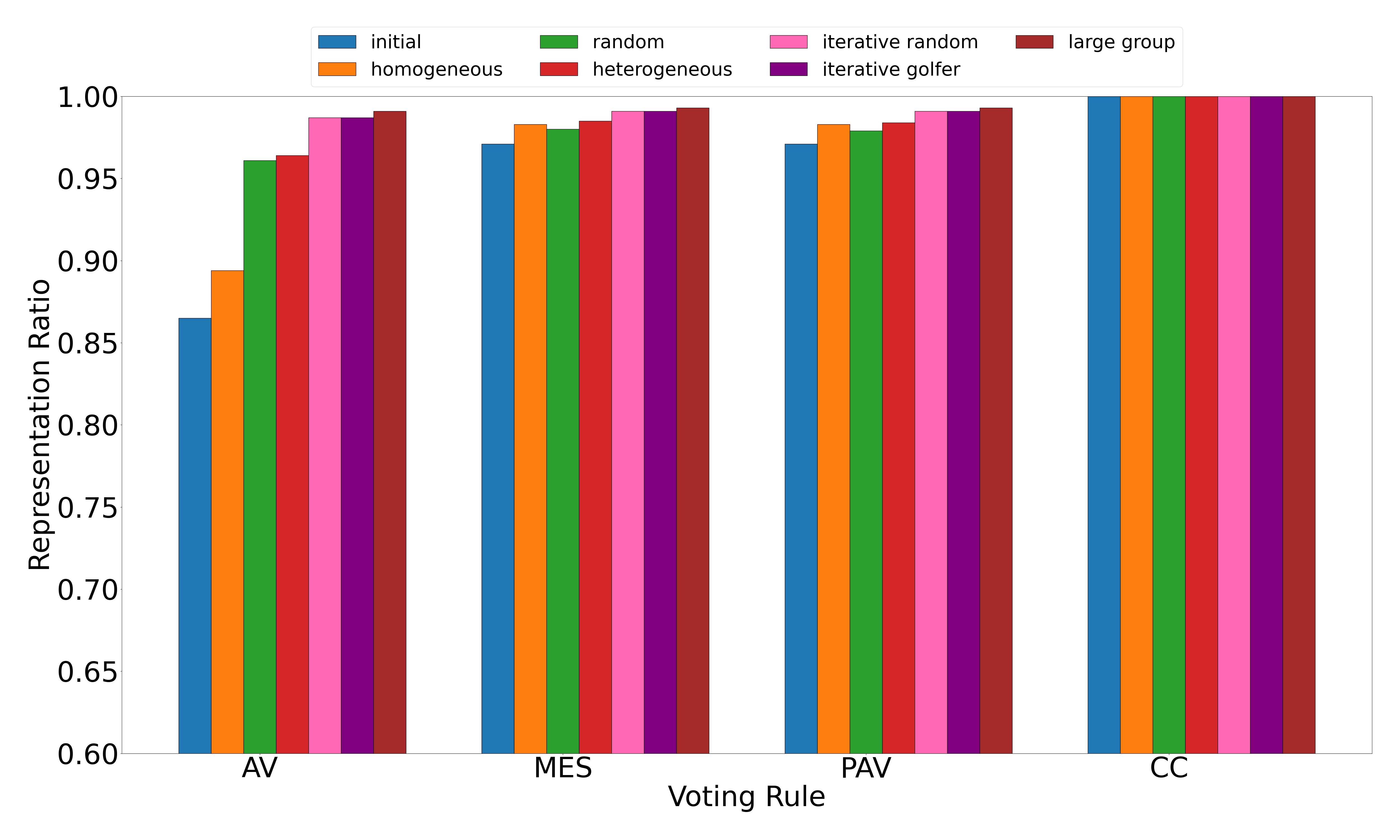}
    \caption{Representation ratio across deliberation mechanisms.}
    \label{fig:representation_ratio}
\end{figure}
\noindent\textbf{Representation Ratio: }
Figure \ref{fig:representation_ratio} shows the average representation ratio of the voting rules across different deliberation mechanisms. Since CC optimizes for diversity by design, $RR(CC)=1.0$.  
Under no deliberation, AV has the lowest representation ratio. Since AV simply picks the candidates with the highest approval scores, it does not care about minority preferences. The proportional rules (MES and PAV), however, achieve much higher representation as they are designed to  maintain a balance between welfare and diversity.

The effect of deliberation is more pronounced here compared to the utilitarian ratio results, particularly for AV.  Within the single-round mechanisms, \textit{homogeneous} achieves a slight improvement over the \textit{initial} setup for all rules. However, specifically for AV, both \textit{heterogeneous} and \textit{random} achieve much higher representation over both \textit{initial} and \textit{homogeneous} setups. Again, both iterative mechanisms achieve further improvements compared to the single-round setups and almost match the \textit{large group} benchmark.

\begin{figure}[t]
    \centering
    \includegraphics[width=\linewidth]{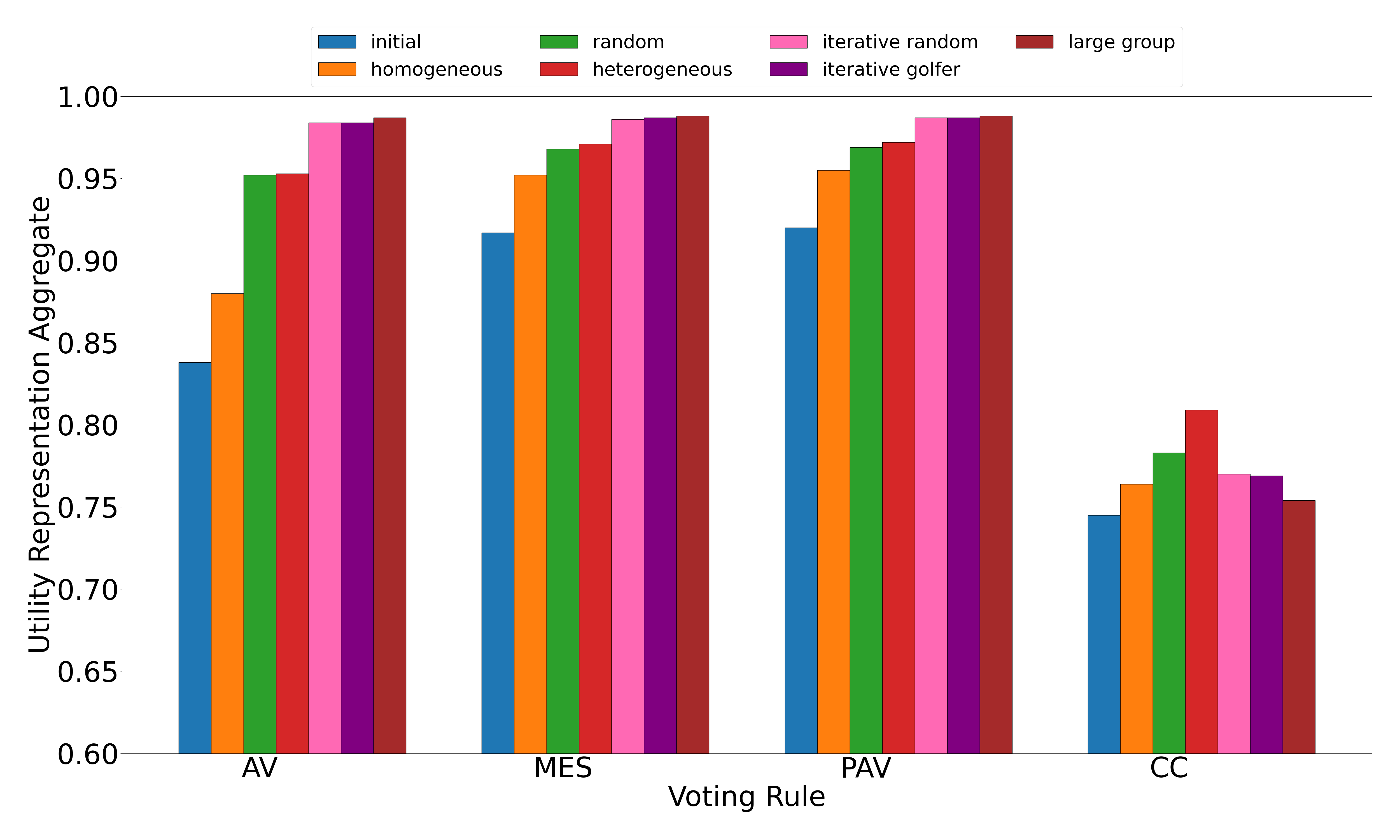}
    \caption{Utility-Representation aggregate score across deliberation mechanisms.}
    \label{fig:aggregate_score}
\end{figure}
\noindent\textbf{Utility-Representation Aggregate Score: }
Figure \ref{fig:aggregate_score} shows the average results for this objective. 
Under no deliberation (\textit{initial} baseline), we see that the proportional rules (MES and PAV) perform the best, followed by AV, and then CC. This is consistent with earlier findings since the proportional rules are designed with this goal in mind. Both AV and CC perform poorly on this metric, pre-deliberation, since they do well on either welfare (AV) or representation (CC) but not both.

We observe a positive effect from deliberation and achieve a significant performance improvement over the \textit{initial} baseline. Within the single-round mechanisms, \textit{heterogeneous} and \textit{random} perform  similarly (except for CC where \textit{heterogeneous} is better) and outperform the \textit{homogeneous} setup. Iterative deliberation leads to further improvement as both iterative methods match the performance of the \textit{large group}.


\begin{table}[t]
\centering
\begin{tabular}{|p{4cm}|c|c|c|c|}
\hline
\centering
\multirow{2}{4cm}{\textbf{Deliberation Strategy}} & \multicolumn{2}{c|}{\textbf{EJR\%}} &\multicolumn{2}{c|}{\textbf{PJR\%}}\\
\cline{2-5}
& \textbf{AV} & \textbf{CC} & \textbf{AV} & \textbf{CC}\\
\hline
Initial (no deliberation)  & 99.5   & 62.5 & 99.5  & 73.4    \\
Homogeneous      & 96.4    & 69.9  & 96.4    & 75.1     \\
Random           & 100      & 81.9  & 100 & 85.6            \\
Heterogeneous    & 100      & 92.7  & 100 & 94.0                \\
Iterative Random & 100      & 31.4  & 100 & 53.6                  \\
Iterative Golfer & 100      & 29.9  & 100 & 51.2                 \\
Large Group      & 100      & 6.10   & 100 & 23.4                 \\
\hline
\end{tabular}
\caption{EJR- and PJR-satisfaction  (AV and CC).}
\label{table:ejr}
\end{table}

\noindent\textbf{EJR and PJR Satisfaction: }
Table \ref{table:ejr} shows the percentage of EJR- and PJR-satisfying committees returned by AV and CC. We focus only on AV and CC since the proportional rules MES and PAV guarantee EJR. Even under no deliberation (\textit{initial}), AV satisfies EJR in almost all profiles, which further improves to perfect satisfaction with deliberation (except \textit{homogeneous}). This is interesting since AV is not guaranteed to satisfy EJR.\footnote{Since the minority and majority agents have highly correlated approval sets, $T$-cohesive  groups may exist only for a small set of minority- and majority-supported candidates, thereby making the EJR requirement easy to satisfy. Furthermore, previous research \cite{welfarevsrep,bredereck2019experimental} shows that under many natural preference distributions (generated elections), there are many EJR-satisfying committees.}
EJR and PJR satisfaction for CC also improves if single-round deliberation is supported, with \textit{heterogeneous} achieving the best result.  Iterative deliberation, however, does not perform well. We believe that this arises due to CC's strong focus on representation (see Appendix \ref{appendix:iter_cc}).

\section{Discussion}
\label{sec:discussion}

Deliberation changes the quality of the outcomes produced by different multi-winner voting rules. We explore these observations in more detail.

\noindent{\bf Single-Round Deliberation:}
Even a single round of deliberation improved outcomes across all voting rules and all objectives. However, the choice of the deliberation structure was also important.
For all objectives, \textit{random} and \textit{heterogeneous} consistently outperformed \textit{homogeneous}.
We hypothesize that this improvement was due to these deliberation strategies maximizing exposure to diverse opinions.
Under \textit{homogeneous} deliberation, the population sub-groups become more inwardly focused, leading to the formation of distinct \textit{T}-cohesive groups. This was particularly problematic when used with AV, which picks candidates with the highest approval support and fails to `fairly' represent the cohesive minority agents in some cases, thereby failing EJR (Table \ref{table:ejr}). By allowing majority and minority agents to interact, there was an opportunity for minority agents to influence the majority population.  This translated to higher welfare, representation, and proportionality guarantees (Figure \ref{fig:aggregate_score} and Table \ref{table:ejr}).


\noindent{\bf Iterative Deliberation:}
In comparison to single-round methods, iterative deliberation further supports consensus (Figure~\ref{fig:variance}) and improves all objectives for most voting rules. Furthermore, there was no statistical difference between the  \emph{iterative golfer} and \emph{iterative random} methods. We view this as a positive result with practical design implications. While care does need to be taken in determining group sizes, a simple, computationally inexpensive mechanism is as effective as one that is more complex.

The exception to the observation is the CC rule. CC’s strong focus on representation and coverage makes it unsuitable for deliberation methods that drive higher degrees of consensus (such as iterative methods and \textit{large group}) since it fails to represent  population groups proportionally. We refer the interested reader to Appendix~\ref{appendix:iter_cc}.


\begin{table}[t]
\centering
\begin{tabular}{|l|l|}
\hline
\centering
\textbf{Deliberation Strategy}   & \textbf{Minority Opinion Preservation} \\
\hline
Initial (no deliberation)          & 0                                \\
Homogeneous      & 0.20                            \\
Random           & 0.30                            \\
Heterogeneous    & 0.48                            \\
Iterative Random & 0.65                             \\
Iterative Golfer & 0.66                             \\
Large Group      & 0.92                           \\
\hline
\end{tabular}
\caption{Minority Opinion Preservation: average number of initial (pre-deliberation) \textit{minority-supported} candidates selected by AV in the final committee after deliberation.}
\label{table:minority}
\end{table}



\noindent{\bf Minority Opinion Preservation:}
While we have been extolling deliberation, there are caveats. 
In particular, it is important to ensure that deliberation processes are inclusive and encourage minority participation~\cite{gherghina_2021_deliberation_minorities}. Care must be taken to ensure that when moving toward consensus, initial minority preferences are not ignored. While consensus would imply better voting outcomes, it could come at the cost of ignoring minority opinions. We measure whether this is a concern in our experiments by studying whether minority-supported candidates were selected by AV under different deliberation mechanisms.\footnote{This is not a concern for other rules since they are designed to achieve proportionality (MES and PAV) or representation (CC).} 

A candidate is either \emph{minority-supported} or \emph{majority-supported} based on the \emph{initial} approval profile. We say that a candidate $c$ is \emph{minority-supported} if (pre-deliberation) the fraction of minority voters who include $c$ in their approval ballot is greater than the fraction of majority voters who include $c$ in their approval ballot. 
Table \ref{table:minority} reports the average number of pre-deliberation (initial) \textit{minority-supported} candidates selected by AV (post-deliberation) across deliberation strategies. This serves as an indicator of whether minority preferences are \textit{preserved}.

In the \textit{initial} setup (no deliberation), AV does not elect any \textit{minority-supported} candidates.  However, this improves  as agents interact and deliberate with the broader population. Note that since the minority agents have similar preferences, and they constitute 20\% of the population in our setup, a proportional committee would represent them with 1 (out of 5) candidate. As seen in Table \ref{table:minority}, the \textit{large group} setup comes close to the ideal outcome on average. Thus, with  deliberation, AV  can \textit{preserve} and represent minority preferences.

\begin{table}[t]
\centering
\begin{tabular}{|l|p{2cm}|p{2cm}|} 
 \hline 
 \textbf{Approval Voting} & \textbf{MES} (initial) \scriptsize{(0.917)} & \textbf{PAV} (initial) \scriptsize{(0.92)}\\
 \hline
 Initial \scriptsize{(0.838)} & 0.913 & 0.910 \\
 Homogeneous \scriptsize{(0.88)} & 0.959 & 0.956 \\
 Random \scriptsize{(0.952)} & \textbf{1.038} & \textbf{1.034} \\
 Heterogeneous \scriptsize{(0.953)} & \textbf{1.039} & \textbf{1.035} \\
 Iterative Random \scriptsize{(0.984)} & \textbf{1.073} & \textbf{1.069} \\
 Iterative Golfer \scriptsize{(0.984)} & \textbf{1.073} & \textbf{1.069} \\
 \hline
\end{tabular}
\caption{Average utility-representation aggregate score obtained by AV under different deliberation setups in comparison to the proportional rules under no deliberation. 
}
\label{table:comparison}
\end{table}

\noindent{\bf ``Simple" vs. ``Complex" Voting Rules:}
We argue that the ``complexity'' of a voting rule can be measured along three axes. First, one can ask about the computational complexity of computing a winning outcome or committee (e.g. PAV is known to be NP-hard~\cite{aziz2014computational}, whereas AV is polynomial). Second, there is growing work in better understanding the ramifications of ballot design and voting rules on the cognitive load of voters~\cite{benade_working_paper,benade2021preference}. 
Finally, there is value in using simple explainable voting rules. Explainability engenders trust in the system (which in turn may impact engagement in participatory democratic processes).

While two of these dimensions are, somewhat subjective, we argue that AV can be viewed as being \emph{simple} across all three, whereas PAV and MES are \emph{complex} along at least one dimension. Our hypothesis is that \emph{simple} rules coupled with deliberation processes can do as well as more \emph{complex} voting rules. To this end, we compare AV with deliberation to MES and PAV without deliberation, using the utility-representation aggregate score ($URagg(R)$) as our measure (Table~\ref{table:comparison}). Values greater than 1.0 indicate that AV with the corresponding deliberation mechanism achieves a better $URagg$ score than MES/PAV without deliberation. These findings support our argument that one does not necessarily have to use ``complex’’ rules as ``simple’’ rules coupled with effective deliberation strategies can be as effective.



\section{Conclusion}\label{sec:conclusion}

We presented an empirical study of the relationship between deliberation and voting rules in approval-based multi-winner elections. Deliberation generally improves voting outcomes with respect to welfare, representation, and proportionality guarantees. Effectively designed mechanisms that increase exposure to diverse groups and opinions enhance the quality of deliberation, protect minority preferences, and in turn, achieve better outcomes. Importantly, we show that in the presence of effective deliberation, `simpler' voting rules such as AV  can be as powerful as more `complex' rules without deliberation. Our hope is that our findings can further support the design of effective citizen-focused democratic processes.

\clearpage


\bibliographystyle{acm}
\bibliography{main}

\clearpage

\appendix
\section*{Appendix}
\section{Opinion Dynamics Models}
\label{appendix:op_dyn}
We discuss two well-established models from opinion dynamics --- DeGroot's classical model~\cite{DeGroot_1974} and Hegselman and Krause's Bounded Confidence (BC) model~\cite{hegselmann_2002}. 

According to DeGroot's classical model~\cite{DeGroot_1974}, an agent's updated opinion is simply the weighted sum of opinions from various sources (itself included). The weights were static, and could be different for different agents. So, for two agents $x$ and $y$, $x$ updates its opinion as:
\begin{equation}
    x(t+1) = w_{xx} x(t) + w_{xy} y(t) 
\end{equation}
where $x(t)$ denotes the opinion of agent $x$ at time $t$, $w_{xx}$ and $w_{xy}$ denote $x$'s weights on its own opinion and $y$'s opinion, respectively. Note that the weights should sum up to 1, and therefore, $w_{xy} = 1 - w_{xx}$. 

Later, there was the Bounded Confidence (BC) model~\cite{hegselmann_2002} which introduced a global confidence level $\Delta$. In the original paper, agents were on a network, and agents updated their opinions based on opinions of their neighbors. In the BC model, an agent $x$ considered a neighbor's ($y$) opinion only if the neighbor's opinion was within $x$'s confidence interval $[x(t) - \Delta, x(t) + \Delta]$. In the initial version, there were no distinct weights and all opinions within the confidence interval were weighted equally. When simplified for just two agents $x$ and $y$, the opinion update for $x$ is given by:
\begin{equation}
\resizebox{\linewidth}{!}{$
    x(t+1) = 
    \begin{cases}
        1/2 (x(t) + y(t), & \text{ if } y(t) \in  [x(t) - \Delta, x(t) + \Delta] \\
        x(t), & \text{ otherwise }
    \end{cases}
    $}
\end{equation}

The BC model captures the idea of confirmation bias, and BC and its several modified versions have largely remained popular till date in the field of opinion dynamics.

\section{Iterative Golfer}
\label{appendix:iter_golfer}
\emph{Iterative golfer} strategy is a weaker version of the popular social golfer problem~\cite{harvey_2002,liu_2019} in combinatorial optimization. 

\begin{quote}
    \emph{Social golfer problem}: $n$ golfers must be repeatedly assigned to $g$ groups of size $s$. Find the maximum number of rounds (and the corresponding schedule) such that no two golfers play in the same group more than once. 
\end{quote}

Social golfer problem maximizes the number of rounds with a hard constraint that no two golfers should meet again. The iterative golfer strategy is a weaker version of this where we fix the number of rounds $R$, and minimize the number of occurrences where any pair of agents meet more than once. Given some group assignment  $G^r = \{G_1^r,G_2^r, \hdots, G_g^r\}$ at round $r$, we introduce a cost given by:

\begin{equation}
    cost(G^r) = \sum_{G_x \in G^r} \sum_{a,b \in G_x} f^2(a,b)
    \label{eq:golfer}
\end{equation}
where $f(a,b)$ is the number of times $a$ and $b$ have been in the same group in the previous rounds $G^1$ through $G^{r-1}$. The number of prior meetings is squared to ensure an even number of conflicts among all possible pairings (as opposed to one specific pair meeting repeatedly). We use an existing approximate solution~\cite{buchanan_2017} that creates group assignment for each round such that the cost given by~\eqref{eq:golfer} is minimized. The iterative golfer can thus be seen as a more efficient strategy than iterative random if the objective is to ensure each agent has the highest possible exposure to others' preferences.

\section{Voter Satisfaction}
\label{appendix:voter_satisfaction}
In Section~\ref{sec:objectives} we introduced a number of objectives on which we compare different voting rules and deliberation processes. 
Another useful objective is \emph{voter satisfaction}, which measures the average number of candidates approved by a voter. 

\noindent\textbf{Voter Satisfaction Score:} Given $W_R=R(A,k)$, the voter satisfaction is measured as the average number of candidates approved by a voter in $W$:
\begin{align}
    VS(R)=\frac{\sum_{i\in N} |A_i\cap W_R| }{|N|}.
\end{align}

\begin{figure}[]
    \centering
    \includegraphics[width=\linewidth]{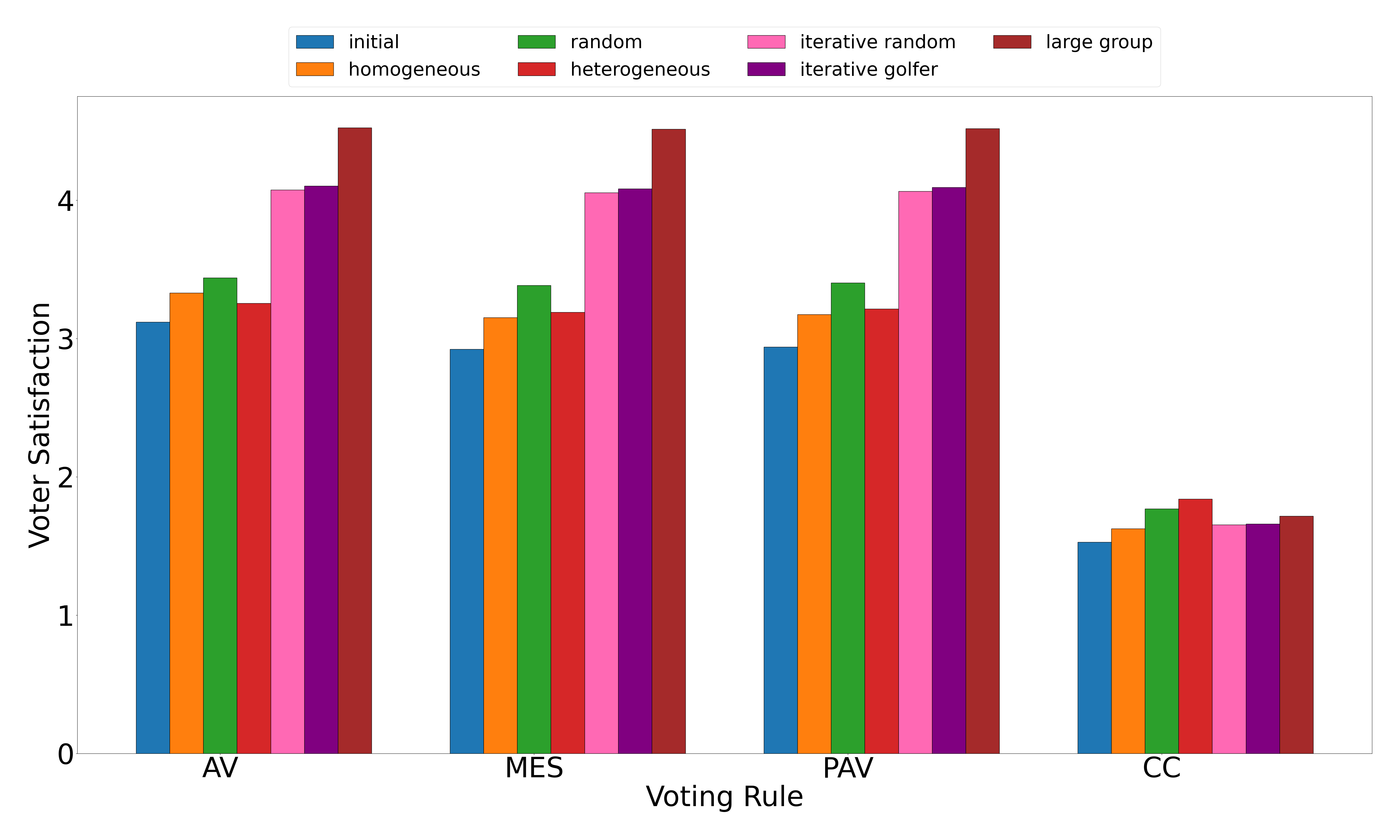}
    \caption{Voter satisfaction achieved by the voting rules across deliberation mechanisms}
    \label{fig:voter_satisfaction}
\end{figure}

Figure \ref{fig:voter_satisfaction} shows the average voter satisfaction obtained by the voting rules across different deliberation setups.

AV is expected to achieve the highest satisfaction since it picks candidates with the highest support, i.e. the average number of candidates approved by a voter will be high. MES and PAV achieve comparable scores, just slightly lower than AV. Finally, CC achieves the lowest satisfaction of all rules. In an attempt to maximize voter coverage, CC might choose winning candidates that represent few voters, and as a result, have low approval scores. Due to this, it maximizes diversity but achieves low voter satisfaction. 

Compared to the \textit{initial} baseline, we observe an improvement in satisfaction scores under all deliberation mechanisms. In general, all single round deliberation setups achieve comparable performance, with the exception of \textit{random} performing the best in some cases. Moving on to the iterative methods, we notice a further increase in satisfaction scores for all rules except CC. While both iterative setups perform similarly and improve over the \textit{initial} baseline, they are still outperformed by the \textit{large group} benchmark.

\section{Iterative deliberation with CC}
\label{appendix:iter_cc}
\begin{figure}[]
    \centering
    \includegraphics[width=\linewidth]{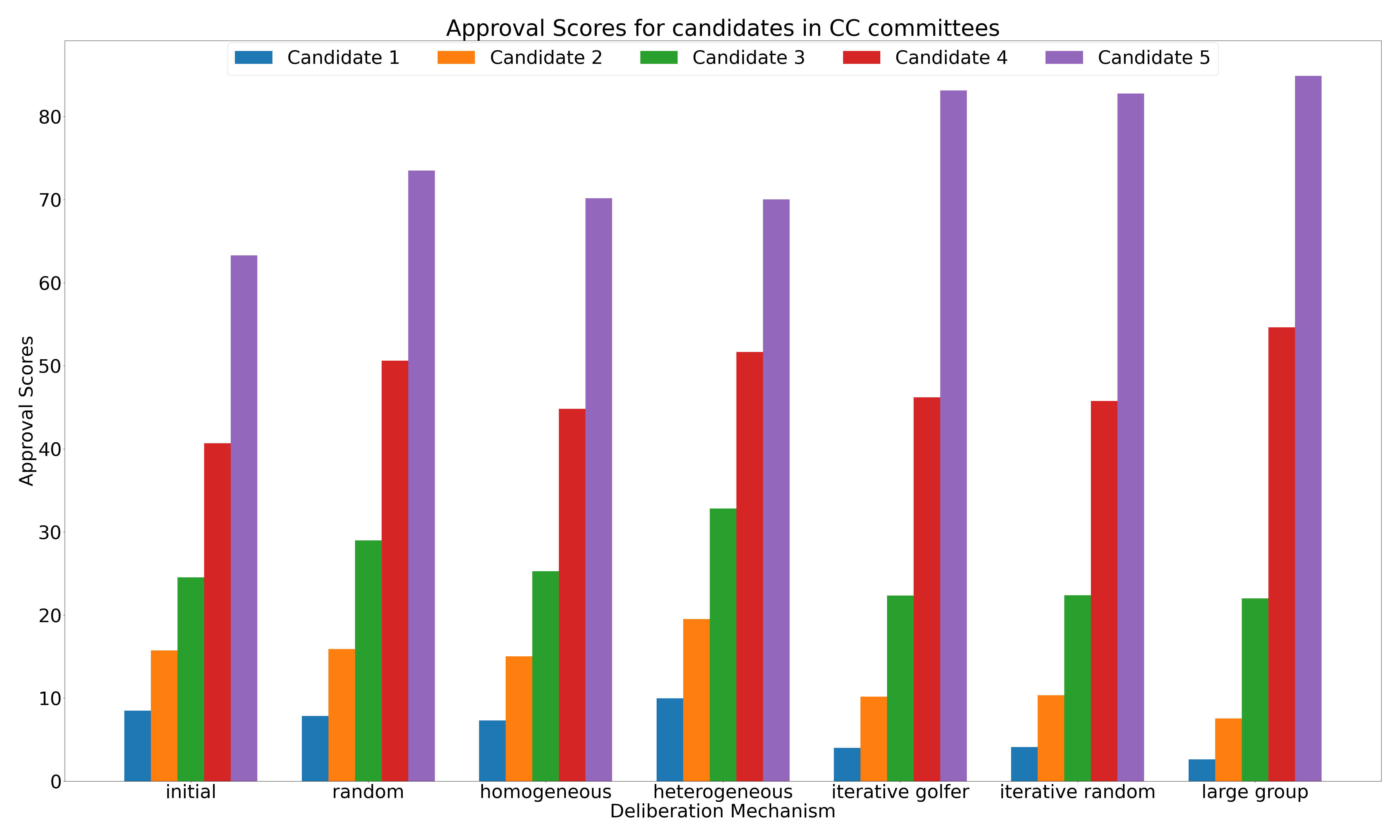}
    \caption{Average approval scores obtained by the 5 candidates in the winning committee chosen by CC across different deliberation mechanisms.}
    \label{fig:cc_approvals}
\end{figure}

Here, we explain the odd drop in performance observed by CC in iterative deliberation and the \textit{large group} setting (see Figures \ref{fig:utilitarian_ratio}, \ref{fig:aggregate_score}, \ref{fig:voter_satisfaction} and Table \ref{table:ejr}). Refer to Figure \ref{fig:cc_approvals} for the average approval scores obtained by the winning candidates in the committees chosen by CC. The candidates (1 to 5) are ranked in increasing order of the number of approval votes they get (5 is highest). 

We  clearly observe that as we move from single round deliberation mechanisms to iterative methods (and \textit{large group}), the approval votes for the highest supported candidate (5) increase and the same for the lowest supported candidate (1) decrease. For the iterative methods, approximately $80\%$ of the agent population approves candidate\_5 ($\approx 90\%$ for \textit{large group}). This also reinforces the fact that iterative deliberation approaches consensus, as a major proportion of voters approve a single candidate. Accordingly, CC is able to represent approximately $80\%$ of the voters with just one candidate. Since CC only cares about maximizing voter coverage, it chooses the rest of the candidates to represent the remaining voters. This leads to sub-optimal outcomes since instead of representing the population groups proportionally, CC optimizes for coverage and chooses candidates that might have very little support. This can be seen in Figure \ref{fig:cc_approvals} as candidate\_1 for the iterative methods and \textit{large group} has less than $5\%$ support. As a result, the almost 80\% of the voter population that possibly gets only one representative in the final CC committee might be a cohesive voter group and thus, deserves more candidates for a fair and proportional outcome.

In conclusion, we see that with deliberation mechanisms that move towards consensus, CC exhibits a drop in welfare and proportionality guarantees since it is focused on maximizing representation. In general, other voting rules provide better overall performance than CC. However, if CC should ever be used with deliberation, we must pick an appropriate deliberation setup (single round) for the optimal outcome. This further shows that deliberation is not trivial and must be structured appropriately to obtain the best results.

\section{Inter-group Ballot Disagreement}
\label{appendix:ballot_disagreement}

\begin{figure}[t]
    \centering
    \includegraphics[width=\linewidth]{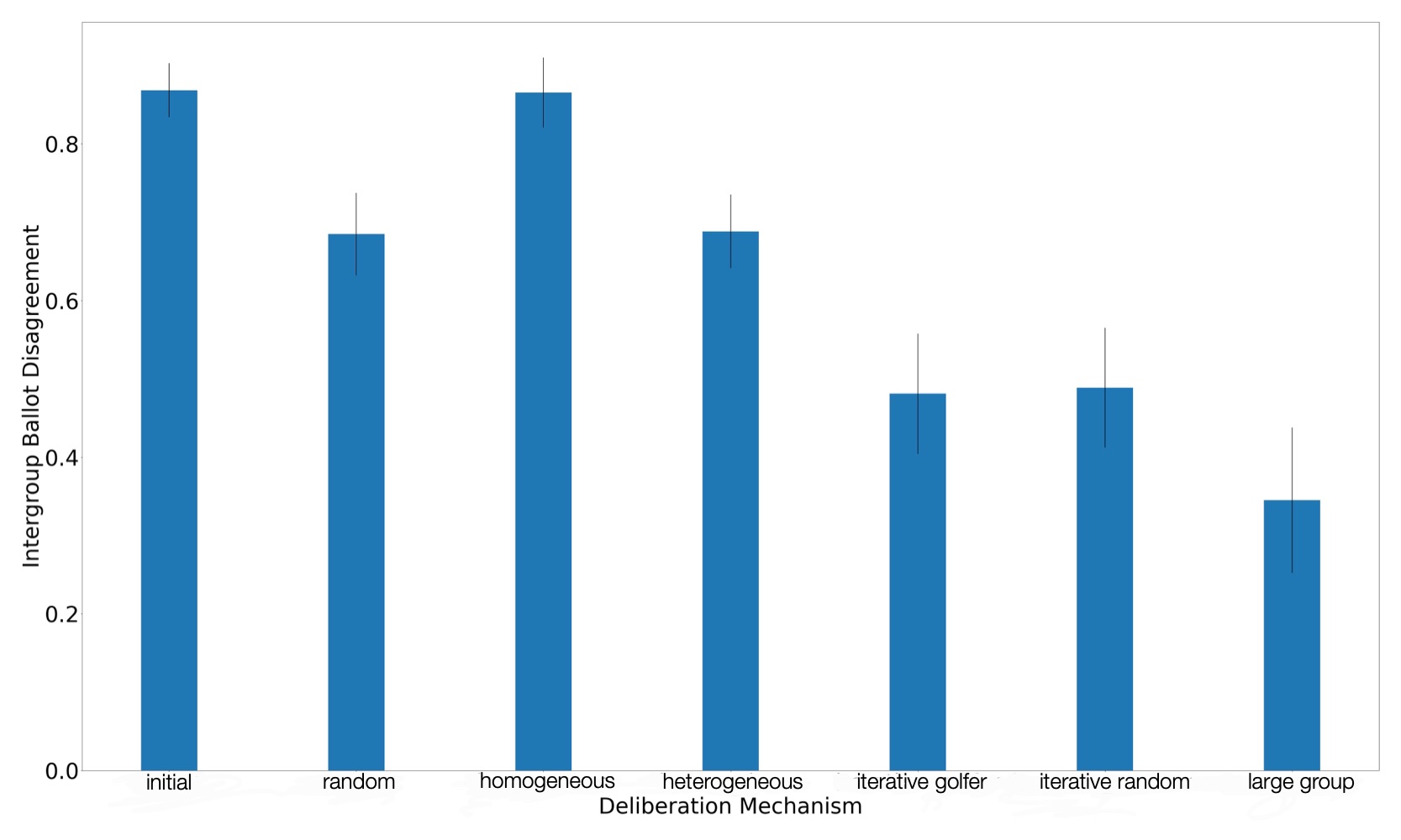}
    \caption{Inter-group Ballot Disagreement}
    \label{fig:ballot_disagreement}
\end{figure}

In Figure \ref{fig:variance} we introduce a measure of consensus in the population as the average variance in agents' utilities and show that deliberation reduces disagreement amongst agents. To complement this analysis and further understand the impact of deliberation on agents' preferences, we introduce another metric that computes the disagreement between the majority and minority voters based on their ballots. In particular, given two approval ballots $A_{min}$ and $A_{maj}$ belonging to a minority voter and a majority voter, respectively, the disagreement score is computed as: $$1 - (|A_{min} \cap A_{maj}|/\min(|A_{min}|, |A_{maj}|))$$ A maximum disagreement score of 1 means the approval ballots are disjoint, \textit{i.e.} the voters do not approve any candidates in common. This score is computed for every majority-minority voter pair in the population and the average results are reported in Figure \ref{fig:ballot_disagreement}.

We observe a similar trend here as well (as seen in Figure \ref{fig:variance}). Deliberation significantly reduces disagreement between the two population groups and moves the overall population toward consensus. This positive effect is stronger in deliberation methods that increase exposure to more, diverse agents (\textit{i.e.} the iterative versions and \textit{large group}). 

\end{document}